\documentstyle[12pt]{article}
\begin{document}
\def\a{\alpha}\def\b{\beta}\def\g{\gamma}\def\d{\delta}\def\e{\epsilon }
\def\k{\kappa}\def\l{\lambda}\def\L{\Lambda}\def\s{\sigma}\def\S{\Sigma}
\def\Th{\Theta}\def\th{\theta}\def\om{\omega}\def\Om{\Omega}\def\G{\Gamma}
\def\y{\vartheta}\def\m{\mu}\def\n{\nu}
\def\ws{worldsheet}
\def\susy{supersymmetry}
\def\ts{target superspace}
\def\ks{$\k$--symmetry}
\newcommand{\plabel}{\label}
\renewcommand\baselinestretch{1.5}
\newcommand{\nn}{\nonumber\\}\newcommand{\p}[1]{(\ref{#1})}
\renewcommand{\thefootnote}{\fnsymbol{footnote}}
\thispagestyle{empty}
\begin{flushright}
Preprint DFPD 00/TH/15\\
hep--th/0003169
\end{flushright}

\bigskip
\begin{center}
{\Large\bf The $type ~IIA$ NS5--Brane}

\vspace{1.5cm}
{\bf Igor Bandos, Alexei Nurmagambetov}

\vspace{0.2cm}
{\small\it Institute for Theoretical Physics\\
NSC Kharkov Institute of Physics and Technology\\
Akademicheskaya 1, \\ UA-61108, Kharkov,  Ukraine\\
bandos@kipt.kharkov.ua, ajn@kipt.kharkov.ua}

\vspace{1.0cm} {\bf Dmitri Sorokin} \footnote{ On leave from {\it
Institute for Theoretical Physics, NSC Kharkov Institute of
Physics and Technology, Kharkov, 61108, Ukraine.}}

\vspace{0.2cm}
{\small\it
INFN, Sezione di Padova, Via F. Marzolo, 8\\
35131 Padova, Italia\\
dmitri.sorokin@pd.infn.it}

\vspace{0.3cm}
{\bf Abstract}
\end{center}
The kappa--invariant worldvolume action for the NS5--brane in a
D=10 type IIA supergravity background is obtained by carrying out
the dimensional reduction of the M5--brane action.

\bigskip

\vspace{3cm}

\newpage
\renewcommand{\thefootnote}{\arabic{footnote}}

\section{Introduction}

The M5-brane plays an important role in studying properties of
M-theory \cite{M}, the theory of strings and associated field theories.
For instance, many physically important multibrane
configurations, realized to be relevant to a brane description of
non-Abelian gauge theories \cite{witten} and a brane-world scenario
\cite{brworld}, can be considered as a specific compactification of
a single D=11 M5-brane down to lower dimensions
(with or without its subsequent T-dualization).

A direct dimensional reduction of D=11 space-time with an M5-brane down
to ten-dimensional space-time produces a so-called NS5-brane
of type IIA supergravity which has been intensively studied
in relation to six-dimensional gauge theories
\cite{6D} and ``little string theories'' \cite{lst}
\footnote{The double dimensional reduction of the M5-brane
action \cite{5br,M5,M51}  is well
known  to result in a type IIA D=10 Dirichlet 4-brane. It has also been
shown  \cite{Berman,3br} how by reducing the M5-brane action one may
arrive at a duality-symmetric D3-brane action.}.

The verification of the quantum consistency of M-theory
requires, in particular, finding a mechanism of anomaly cancellation
in the presence of M5-branes. It has been shown that
the anomaly problem has a natural solution in the case of $D=10$ NS5-branes
\cite{anomaly}, while in the case of the D=11 M5-brane the situation
is much more subtle and requires additional study
\cite{anomaly,bonora,moore,wit}.
Mechanisms for the M5-brane anomaly cancellation proposed recently in
\cite{bonora} include as an important feature the reduction of the
structure group $SO(5)$ of the normal bundle of the M5-brane down to its
$SO(4)$ subgroup. Such a reduction implies an existence of a covariantly
constant vector field and,
therefore, looks very much as a dimensional reduction (to
be more precise, the dimensional reduction is a particular case of such an
`M5-brane framing' \cite{bonora})
\footnote{In contrast to \cite{bonora} the analysis of ref. \cite{moore}
is based on the assumption that a full understanding of anomaly
cancellation requires keeping the full $SO(5)$. We are thankful to Jeff
Harvey for clarifying this difference in the approaches.}.
These facts provide us with a motivation to study in more detail
the dynamical and symmetry properties of the NSIIA five-brane
by constructing a full worldvolume action describing its dynamics in
a type IIA D=10 supergravity background.

By now the action for the NS5--brane has been constructed up to
a second order in the field strength of a two--rank self--dual
worldvolume gauge field
of the five--brane and only in a background of the bosonic sector of IIA
D=10 supergravity \cite{BLO,lozano}.

The aim of this paper is to get a full, nonlinear and $\kappa$--symmetric,
NS5--brane action in a curved IIA D=10 target superspace by carrying out
the direct dimensional reduction of the D=11 M5--brane action
\cite{5br,M5,M51},
and thus filling in a gap in the list of worldvolume actions for
supersymmetric extended objects found in string theory.

The fact that the NS5--brane can be regarded as an M5--brane propagating
in a dimensionally reduced D=11 supergravity background
substantially  simplifies the analysis of the NS5--brane model,
in particular,
allowing one to derive its symmetries and dynamical properties
directly from those of the  M5--brane.

For instance, the physical field content of the IIA D=10 NS5--brane
is the same as of the M5--brane.
The bosonic sector consists of three degrees of freedom
corresponding to the two--rank self--dual worldvolume field and five
worldvolume scalars. In the case of the M5--brane the five scalar fields
describe its oscillations in a D=11 background in the directions
transversal to the M5--brane worldvolume, while in the case of the
NS5--brane four scalar fields correspond to transversal oscillations in a
D=10 background, and the fifth scalar field (corresponding to the
compactified dimension of the D=11 space) `decouples' and becomes a
`purely' worldvolume field. This results in the abovementioned reduction
of the M5-brane normal bundle structure group $SO(5)$ down to $SO(4)$.
For both five--branes eight fermionic fields can be associated with brane
`oscillations' in Grassmann directions of corresponding target
superspaces.

To get the action describing the dynamics of the physical modes of the
NS5--brane as a dimensionally reduced M5--brane action we first
briefly remind the structure and properties of the latter.

In Sections 2--5 we consider bosonic M5-- and NS5--branes and in
Section 6 we describe the full target--superspace covariant and
$\kappa$--invariant NS5--brane action.

\section{The M5-brane action}

In the absence of interactions with antisymmetric tensor fields of $D=11$
supergravity the action for the bosonic sector of the M5-brane
has the following form \cite{5br}
\begin{equation}\plabel{ac}
S=-\int\,d^6\xi\,\left[\sqrt{-\det(\hat{g}_{mn}+i {\hat{H}}^*_{mn})}
+{{\sqrt{-\hat{g}}}\over 4{\sqrt{-\widehat{\partial a\partial a}}}}
\hat{H}^{*mn}{H}_{mnr}\partial^{r} a\right]
\end{equation}
where
$$
m,n,...=0,\dots,5;
$$
are vector indices of $d=6$ worldvolume coordinates $\xi^m$,
$$
{\underline{\hat{m}}},{\underline{\hat{n}}},...=0,\dots,10,
$$
are vector indices of $D=11$ target space coordinates
$\hat X^{\underline{\hat{m}}}$
\begin{equation}\plabel{511}
\hat{g}_{mn}=\partial_m \hat{X}^{\underline{\hat{m}}}
\hat{g}^{(11)}_{\underline{\hat{m}\hat{n}}}
\partial_n \hat{X}^{\underline{\hat{n}}},
\end{equation}
is the worldvolume metric induced by embedding the five--brane into
a $D=11$ gravity background with a metric
$\hat{g}^{(11)}_{\underline{\hat{m}\hat{n}}}(X)$
(we use the `almost minus' Minkowski signature  $(+-\dots-)$),\\
${H}_{mnl}(\xi)=3\partial_{[m}b_{nl]}$ is the field strength of the
worldvolume antisymmetric tensor field ${b}_{mn}(\xi)$,
\begin{equation}\plabel{h}
{\hat{H}}^*_{mn}\equiv
{1\over \sqrt{-\widehat{\partial a \partial a}}}
\hat{H}^{*}_{mnr}\partial^{r} a,
\qquad
\hat{H}^{*mnl}={1\over
{3!\sqrt{-\hat{g}}}}\epsilon^{mnlrsq} {H}_{rsq},
\end{equation}
$a(\xi)$ is an auxiliary scalar field ensuring the
covariance of the model,  and
\begin{equation}\plabel{sp}
\widehat{\partial a\partial a}\equiv \partial_m a {\hat g}^{mn}\partial_n a
\end{equation}
denotes the scalar product of the $d=6$ vector $\partial_m a$ with
respect to the metric (\ref{511}). In what follows the `hat' over
quantities indicates that they correspond to or induced by the
eleven--dimensional theory.

In addition to the usual gauge symmetry of the $b_2$ field
\begin{equation}\plabel{gb}
\d a(\xi)=0,\ \ \ \d b_{mn}=2\partial_{[m}\varphi_{n]}(\xi),
\end{equation}
the action \p{ac} is invariant under the following transformations
\cite{5br}, \cite{M5}, \cite{M51}
\begin{equation}\plabel{sym}
\d a (\xi)=0,\ \ \ \d b_{mn}=2 \phi_{[m}(\xi) \partial_{n]}a(\xi),
\end{equation}
\begin{equation}\plabel{sym1}
\d a=\varphi(\xi),\ \ \  \d b_{mn}={\d a\over
\sqrt{ -\widehat{\partial a \partial a}}}
[\hat{{\cal H}^*}_{mn} - {H}_{mnp}\hat{g}^{ps}
{ \partial_s a \over \sqrt{-\widehat{\partial a \partial a} } } ],
\end{equation}
where
\begin{equation}\plabel{calv}
\hat{{{\cal H}^*}}_{mn}= - { 2 \over \sqrt{-\hat{g}}}
{\d {\cal{L}}_{DBI}\over \d\hat{{H}}^{*mn}}, \qquad
{\cal{L}}_{DBI} \equiv
\sqrt{-\det(\hat{g}_{mn}+i\hat{{H}}^*_{mn})}.
\end{equation}
Note that at the linearized level, $\hat{{{\cal H}^*}}_{mn}$
defined in (\ref{calv}) reduces to $\hat{{H}}^*_{mn}$.

The symmetries (\ref{sym}) and (\ref{sym1})
are characteristic of the covariant approach \cite{pst}
to the Lagrangian description of duality--symmetric fields.
They ensure the $b_2$ field equation of motion to reduce to a
self--duality condition, as well as the connection with
non-covariant formulations \cite{noncov,mps1}

Let us briefly describe how one derives the symmetries
(\ref{sym}) and (\ref{sym1}) and gets the self--duality condition
\cite{5br}, \cite{pst}.

To this end note that the second term in the action \p{ac}
can be written in terms of differential forms
\begin{equation}\plabel{L1}
\int\, d^6\xi {\cal L}_1 \equiv -\int\, d^6\xi
\sqrt{-\hat{g}}{1\over 4{\sqrt{-\widehat{\partial a \partial a}}}}
\hat{{H}}^{*mn}{H}_{mnr}\partial^{r} a
= -\int_{{\cal M}^6}\, {1 \over 2} \hat{v} \wedge H_3 \wedge
i_{\hat{v}} H_3,
\end{equation}
where
\footnote{In our notation
$  d\xi^{m_1} \wedge \ldots \wedge d\xi^{m_6} =
d^6 \xi \e^{m_1  \ldots m_6}$.}
\begin{equation}\plabel{hatv}
\hat{v}= d\xi^m \hat{v}_m , \qquad
\hat{v}_k\equiv {\partial_k a \over \sqrt{-\widehat{\partial a \partial a}}}
\end{equation}
\begin{equation}\plabel{H3}
H_3 \equiv {1 \over 3!} d\xi^m \wedge d\xi^n \wedge d\xi^l H_{lnm},
\qquad
 i_{\hat{v}}H_3 \equiv {1 \over 2}
d\xi^m \wedge d\xi^n \hat{v}_k ~ \hat{g}^{kl} H_{lnm}.
\qquad
\end{equation}
The variation of the first term in \p{ac} with respect to the gauge field
and the scalar $a(\xi)$
can be written in terms of differential forms as
\begin{equation}\plabel{dLDBI}
\int\, d\xi^6 \d {\cal{L}}_{DBI} \equiv \int\, d\xi^6 \d
\sqrt{-\det(\hat{g}_{mn}+i\hat{{H}}^*_{mn})}
= \int_{{\cal M}^6}\, ~\hat{\cal H}^*_2 \wedge {*}
\d \hat{H}^*_2,
\end{equation}
where 2-forms $\hat{\cal H}^*_2$ and $\hat{H}^*_2$ are constructed
respectively from the tensors \p{calv} and \p{h}
\begin{equation}\plabel{hatG2}
\hat{\cal H}^*_2 \equiv {1 \over 2} d\xi^m \wedge d\xi^n \hat{{\cal H}^*}_{nm},
\qquad
\hat{H}^*_2 = i_{\hat{v}} {*}H_3
\equiv {1 \over 2} d\xi^m \wedge d\xi^n \hat{H}^*_{nm}
\qquad
\end{equation}
and ${*}$ is the Hodge operation in $d=6$ dimensions
\footnote{To have $**=I$ we define
$$
 * \Omega_2 = -{1 \over 2! 4!} d\xi^{m_4} \wedge
 \ldots \wedge d\xi^{m_1} \sqrt{-g}
 \epsilon_{m_1 \ldots m_4 n_1n_2} \Omega^{n_1n_2},
$$
$$
* \Omega_4 =
+ {1 \over 2! 4!} d\xi^{m_2}
\wedge d\xi^{m_1} \sqrt{-g}
 \epsilon_{m_1 m_2 n_1 \ldots n_4} \Omega^{n_1  \ldots  n_4}\equiv
 {1 \over 2! 4!} d\xi_{m_2}
\wedge d\xi_{m_1} {1 \over \sqrt{-g}}
 \epsilon^{m_1 m_2 n_1 \ldots n_4} \Omega_{n_1  \ldots  n_4}
$$}.

Using the identities
\begin{equation}\plabel{Id0}
i_{\hat{v}} \delta \hat{v} =0, \qquad
\Omega_6 \equiv - i_{\hat{v}}\Omega_6 \wedge \hat{v}, \qquad
*i_{\hat{v}} * {H}_3 = H_3 \wedge \hat{v},
\end{equation}
\begin{equation}\plabel{Id1}
 \hat{v} \wedge H_3 \wedge i_{\hat{v}}\d H_3
= \hat{v} \wedge i_{\hat{v}}H_3 \wedge \d H_3 + H_3
\wedge\d H_3, \qquad
\end{equation}
\begin{equation}\plabel{Id2}
\hat{v} \wedge H_3 \wedge i_{\d \hat{v}} H_3 =
\d \hat{v} \wedge *H_3 \wedge i_{\hat{v}} *H_3
=- \d \hat{v}\wedge \hat{v} \wedge i_{\hat{v}} {{*}}H_3
\wedge i_{\hat{v}} {{*}}H_3
\end{equation}
and
\begin{equation}\plabel{relGG2}
\hat{v} \wedge \hat{\cal H}^*_2 \wedge\hat{\cal H}^*_2 = \hat{v} \wedge
\hat {H}^*_2 \wedge \hat{H}^*_2 \quad
\Longleftrightarrow \quad
\e^{abcdef}\hat{{\cal H}^*}_{bc}\hat{{\cal H}^*}_{de}\hat{v}_f
=\e^{abcdef}\hat{H}^*_{bc}\hat{H}^*_{de}\hat{v}_f,
\end{equation}
one can rewrite the variation of the Lagrangian \p{ac} in the form
\begin{equation}\plabel{dL}
\int\, d\xi^6 \d {\cal{L}} \equiv -\int_{{\cal M}^6}
({1 \over 2} H_3 \wedge \d H_3 - da \wedge {\cal F}_2  \wedge \d H_3
- {1\over 2} d\d a \wedge da \wedge {\cal F}_2  \wedge {\cal F}_2),
\end{equation}
where
\begin{equation}\plabel{calF2}
{\cal{F}}_2 \equiv
{1 \over \sqrt{-\widehat{\partial a \partial a}}}
 ({\hat{{\cal H}^*}} - i_{\hat{v}} H_3) =
{1 \over 2} d\xi^m\wedge d\xi^n {\cal{F}}_{nm},
\end{equation}
or
\begin{equation}\plabel{calF2mn}
{\cal{F}}_{mn} \equiv
{1 \over \sqrt{-\widehat{\partial a \partial a}}}
 (\hat{\cal H}^*_{mn} -
H_{mnl}\hat{g}^{lk} {\partial_k a
\over \sqrt{-\widehat{\partial a \partial a}}}).
\end{equation}

Since $H_3=db_2$,  the variation \p{dL} can be written
(up to a total derivative) in the following form
\footnote{We use conventions where external derivative acts from the right:
$$
d\Omega_q = {1 \over q!} d\xi^{m_q} \wedge \ldots \wedge d\xi^{m_1}
\wedge d\xi^{n} \partial_n \Omega_{m_1\ldots m_q}, \qquad
d(\Omega_p \wedge \Omega_q) = \Omega_p \wedge d\Omega_q
+ (-)^q d\Omega_p \wedge \Omega_q.  $$}
\begin{equation}\plabel{dL0}
\int\, d\xi^6 \d {\cal{L}} \equiv - \int_{{\cal M}^6}
d(da \wedge {\cal F}_2)  \wedge (\d b_2-\d a {\cal F}_2),
\end{equation}
 from which the invariance of the action under (\ref{sym}) and (\ref{sym1})
becomes evident.

 From (\ref{dL0}) it also follows that the equation of motion of $b_2$ field is
\begin{equation}\plabel{Beq}
d(da \wedge {\cal F}_2)=0,
\end{equation}
and the equation of motion of $a(x)$ is a consequence of eq. (\ref{Beq}).
It can be shown \cite{pst} that, using the symmetry (\ref{sym}),
the second--order equation (\ref{Beq}) reduces to the first--order
self--duality condition
\begin{equation}\plabel{selfddf}
{\cal{F}}_2 \equiv
{1 \over \sqrt{-\widehat{\partial a \partial a}}}
 ({\hat{{\cal H}^*}} - i_{\hat{v}} H_3) = 0,
\end{equation}
or in components
\begin{equation}\plabel{selfdmn}
 \hat{\cal H}^*_{mn} =
H_{mnl}\hat{g}^{lk} {\partial_k a
\over \sqrt{-\widehat{\partial a \partial a}}}.
\end{equation}

To prove this note that $\hat{H}^{*}_2$ is invariant
under the transformations \p{sym}
$$
  \delta b_2 =
  da \wedge \phi_1
  \equiv
\sqrt{ -\widehat{\partial a \partial a} }~
  \hat{v} \wedge \phi_1
\quad \Rightarrow $$ $$ \delta \hat{H}^{*}_2 \equiv \delta
i_{{\hat v}} (* \hat{H}_3) = \sqrt{ -\widehat{\partial a \partial
a}}~ i_{\hat{v}} (* (\hat{v} \wedge d\phi_1)) \equiv 0.
$$
Hence,
the transformations of the two-form \p{calF2} reduce to
\begin{equation}\plabel{df2}
  \delta {\cal F}_2
=-{1\over \sqrt{ -\widehat{\partial a \partial a} }}~
\delta i_{\hat{v}}H_3
=-
i_{\hat{v}} (\hat{v} \wedge d\phi_1).
\end{equation}
Eq. (\ref{df2}) is simplified when one takes into account that
$i_{\hat{v}} da =
 \sqrt{ -\widehat{\partial a \partial a} } ~
i_{\hat{v}} \hat{v} = -
 \sqrt{ -\widehat{\partial a \partial a} }.
$
Then
$$
\delta {\cal F}_2 =- d\phi_1 +
i_{\hat{v}} d\phi_1 \wedge \hat{v},
$$
and
\begin{equation}\plabel{daf}
\delta (da \wedge  {\cal F}_2) =- da \wedge d\phi_1.
\end{equation}
We now observe that eq. (\ref{daf}) is similar to the
general solution of eq. \p{Beq} for  $da \wedge  {\cal F}_2 $.
This means that the general solution of eq. \p{Beq} can be gauged to zero
with the use of the symmetry \p{sym}, and eq. \p{selfddf} appears just as
a result of such gauge fixing.

Remember that $\hat{\cal H}^*_{mn}$ is defined in (\ref{calv}) and reduces
to  $\hat{H}^*_{mn}=H^*_{mnl}\hat{g}^{lk} {\partial_k a
\over \sqrt{-\widehat{\partial a \partial a}}}$ at the linearized level,
the equation (\ref{selfdmn}) becoming the conventional self--duality condition
$\hat H^*_{lmn}=H_{lmn}$.
Further details on the classical dynamics of the M5--brane the reader may
find in \cite{5br,M5,M51}, \cite{hs2}--\cite{physrep}.

\section{Dimensional reduction of $D=11$ gravity and the NS5-brane action}

The procedure of the direct dimensional reduction assumes a
compactification of some of target--space spatial
dimensions (one in our case), the worldvolume of the
$p$--brane being not compactified.
A standard (string frame) ansatz for the target--space vielbein under the
Kaluza-Klein reduction of one spatial dimension has the following form
$$
E^{\hat{\underline{a}}}= (E^{\underline{a}}, E^{{10}}) \equiv
d\hat{X}^{\hat{\underline{m}}}
{E_{\hat{\underline{m}}}}^{\underline{\hat a}}(\hat{X}),
\qquad \hat{X}^{\hat{\underline{m}}} = (X^{\underline{m}}, y),
\qquad
y=\hat{X}^{{10}},
$$
$$
E^{\underline{a}}= e^{-{1\over 3}\Phi} dX^{\underline{m}}
e_{\underline{m}}^{~\underline{a}}(X), \qquad
E^{{10}}=
e^{{2\over 3}\Phi} (dy-dX^{\underline{m}} A_{\underline{m}})
\equiv e^{{2\over 3}\Phi}{\cal F},
$$
\begin{equation}\plabel{3}
{e_{\underline{\hat m}}}^{\underline{\hat a}}=
\left(\begin{array}{cc} e^{-{1\over 3}\Phi}
{e_{\underline{m}}}^{\underline{a}} & -e^{{2\over
3}\Phi}A_{\underline{m}}\\
0 & e^{{2\over 3}\Phi}
\end{array}\right),
\end{equation}
where $y$ is the coordinate compactified into a torus, and
the reduction means that the background fields,
such as components of (\ref{3}), do not depend on $y$ which is now considered
as an intrinsic scalar field in the 5--brane worldvolume.
$\Phi(X)$ is the dilaton field and $A_{\underline m}(X)$
is the Abelian vector gauge field of $D=10$ IIA supergravity.
The $U(1)$--gauge transformations of $A_{\underline m}(X)$ and $y$ are
\begin{equation}\plabel{u1}
\delta A_{\underline m}(X)=\partial_{\underline m}\varphi^{(0)}(X), \quad
\delta y=\varphi^{(0)}(X).
\end{equation}

This ansatz leads to the following expression for the $D=11$ target space
metric in terms of the $D=10$ metric
$g^{(10)}_{\underline{mn}}(X)=e_{\underline{m}}^{\underline{a}}
e_{\underline{ma}}$, $A_{\underline m}(X)$ and $\Phi(X)$
\begin{equation}\plabel{4}
\hat{g}^{(11)}_{\underline{\hat{m}\hat{n}}}=
\left(\begin{array}{cc} e^{-{2\over 3}\Phi}(g^{(10)}_{\underline{mn}}-e^{2\Phi}
A_{\underline{m}}A_{\underline{n}}) & e^{{4\over
3}\Phi}A_{\underline{m}}\\ e^{{4\over 3}\Phi}A_{\underline{n}} &
-e^{{4\over 3}\Phi} \end{array}\right)
\end{equation}
and, consequently, to the following form of
the six--dimensional induced metric \p{511}
\begin{equation}\plabel{5}
\hat{g}_{mn} = e^{-{2\over3}\Phi}(g_{mn}-e^{2\Phi}
{\cal{F}}_m{\cal{F}}_n).
\end{equation}
In (\ref{5})
\begin{equation}\plabel{g10}
{g}_{mn} =\partial_m X^{\underline{m}}g^{(10)}_{\underline{mn}}(X)
\partial_n X^{\underline{n}}, \qquad \underline{m}=0,\dots,9
\end{equation}
is the six-dimensional metric induced by embedding the 5-brane
worldvolume into the ten--dimensional curved space-time
and
\begin{equation}\plabel{6}
{\cal{F}}_m=\partial_m y- A_m,
\end{equation}
where
$A_m=\partial_m X^{\underline{m}} A_{\underline{m}}(X)$ is the worldvolume
pullback of $A_{\underline{m}}(X)$ and ${\cal{F}}_m$ is the pullback of
the one--form $\cal F$ introduced in \p{3}.

${\cal{F}}_m$ defined in (\ref{6}) can be considered as a field strength
of the worldvolume scalar field $y(\xi)$. It is invariant under the
$U(1)$ gauge transformations (\ref{u1}).

In what follows we will also use an expression for the
inverse worldvolume metric
\begin{equation}\plabel{7}
\hat{g}^{mn}=e^{{2\over3}\Phi}({g}^{mn}+{e^{2\Phi}{\cal{F}}^m{\cal{F}}^n\over
{1-e^{2\Phi}{\cal{F}}^2}}).
\end{equation}

The NS5--brane action follows from the M5-action \p{ac} with
the background metric having a particular form \p{4}
and the coordinate $\hat{X}^{\underline{10}}=y(\xi)$ being
considered as an intrinsic worldvolume scalar field.
To present the explicit form of the NS5--brane action we should rewrite all
its constituents in terms of $D=10$ fields, and to `rescale' worldvolume
fields and their scalar products with respect to the worldvolume induced
metric (\ref{g10}).

For instance, the Hodge duality (\ref{h}) is now redefined with
respect to the metric (\ref{g10})
\begin{equation}\plabel{h*}
\hat{H}^{*mnp}=\sqrt{{{g}\over \hat{g}}}H^{*mnp}, \quad
{H}^{*mnl}={1\over
{3!\sqrt{-{g}}}}\epsilon^{mnlrsd} {H}_{rsd},
\end{equation}
and the M5--brane field strength  $\hat{{H}}^{*mn}$ (\ref{h}) is related to
its NS5--brane  counterpart ${{H}}^{*mn}$ as
\begin{equation}\plabel{h1}
\hat{{H}}^{*mn}
=
{\sqrt{{{g}\over \hat{g}}}\sqrt{{(\partial a)^2}\over
{\widehat{\partial a \partial a}}}}
{{H}}^{*mn}, \qquad
{{H}}^{*mn}=
{1\over {3!\sqrt{-{g}}}}\e^{kmnpqr} {H}_{pqr}
{\partial_k a\over \sqrt{-(\partial a)^2}},
\end{equation}
where the scalar product (\ref{sp}) has also been correspondingly redefined

as
\begin{equation}\plabel{v1}
\widehat{\partial a \partial a} \equiv
\partial_k a
\hat{g}^{ks} \partial_s a
= e^{{2\over 3}\Phi} {\cal N}^2 (\partial a)^2, \quad
(\partial a)^2\equiv \partial_l a ~{g}^{lm} ~\partial_m a,
\end{equation}
with ${\cal N}$ standing for
\begin{equation}\plabel{r}
{\cal N} \equiv \sqrt{1+{{
e^{2\Phi}({\cal F}\partial a)^2}
\over{(\partial a)^2(1-e^{2\Phi}{\cal F}^2)}}}=e^{-{1\over 3}\Phi}\sqrt{
{{\widehat{\partial a \partial a}}\over
(\partial a)^2}}.
\end{equation}
In view of eqs. (\ref{5}), (\ref{h1}), (\ref{v1})  and (\ref{r})
the antisymmetric tensor entering the DBI-like part of the M5--brane action
is reexpressed in terms of ${H}^{*lm}$ as follows
\begin{equation}\plabel{hns2}
\hat{{H}}^*_{mn}=\hat{g}_{ml}\hat{g}_{nk}{\hat H}^{*lk}=
\hat{g}_{ml}\hat{g}_{nk}\sqrt{{{g}\over \hat{g}}}
e^{-{1\over 3}\Phi}{\cal N}^{-1}{H}^{*lk}, \qquad
\hat{g}_{ml} = e^{-{2\over3}\Phi}(g_{ml}-e^{2\Phi}
{\cal F}_m
{\cal F}_l).
\end{equation}
As a result, substituting (\ref{5})--(\ref{hns2}) into the action
(\ref{ac}), we get the action for a bosonic 5--brane coupled to the
metric, the dilaton and the gauge vector field of type IIA $D=10$
supergravity
$$
S=-\int\,d^6\xi\,e^{-2\Phi}\sqrt{-\det({g}_{mn}-e^{2\Phi}{\cal{F}}_m
{\cal{F}}_n)}\sqrt{\det\left({\d_m}^n+i{e^{\Phi}({g}_{mp}-
e^{2\Phi}{\cal{F}}_m
{\cal{F}}_p)\over {{\cal N}\sqrt{\det({\d_m}^n-e^{2\Phi}{\cal{F}}_m
{\cal{F}}^n)}}}{H}^{*np}\right)}
$$
\begin{equation}\plabel{ns2ac}
-{1\over 4} \int d^6\xi\, \sqrt{-{g}}
{1 \over{\cal N}^2}
{H}^{*mn}H_{mnk}\left({g}^{kp}+{e^{2\Phi}{\cal{F}}^k{\cal{F}}^p
\over{1-e^{2\Phi}{\cal{F}}^2}}\right){\partial_p a \over \sqrt{-(\partial
a)^2}}.
\end{equation}

Since the action \p{ns2ac} is nothing but the M5-brane action
for a special choice of the $D=11$ metric (\ref{4}), its variation with respect
to the gauge field $b_2(\xi)$ and the auxiliary scalar $a(\xi)$
has the form of eq. \p{dL0}, and hence \p{ns2ac} is also invariant under
the symmetries \p{sym} and \p{sym1} which, as we have seen, produce
the self-duality condition \p{selfdmn}.

To rewrite the transformations  and the self--duality condition \p{selfdmn}

in the form adapted to the NS5--brane propagating in the $D=10$ background,
let us introduce the NS5 counterpart
of the tensor $\hat{\cal H}^*_{mn}$ \p{calv}
\begin{equation}\plabel{Gmn}
{\cal H}^*_{mn}= -{2\over \sqrt{-{g} }}{\delta {\cal{L}}_{kin.NS5}
\over {\delta{H}^{*mn}}},
\end{equation}
where ${\cal{L}}_{kin.NS5}$ denotes the first (DBI-like) term  in the
action \p{ns2ac}, which is just the DBI--like term
of the M5-action \p{ac} written in the $D=10$ adapted worldvolume frame.
Using \p{h1}, it is easy to find
the relation between $\hat{\cal H}^*_{mn}$ and ${\cal H}^*_{mn}$
\begin{equation}\plabel{GhatG}
{\cal H}^*_{mn}= \hat{{\cal H}^*}_{mn}
\sqrt{(\partial a)^2 \over {\widehat{\partial a\partial a}}}.
\end{equation}
Taking into account eqs.
\p{selfdmn}, \p{7}, \p{v1} and \p{GhatG} we obtain
the following form of the
local worldvolume symmetries
\begin{equation}\plabel{symns2}
\d a=0,\ \ \ \d b_{mn}= -2 \partial_{[m}a ~\phi_{n]}(\xi),
\end{equation}
$$
\d a=\varphi(\xi),
$$
\begin{equation}\plabel{symns21}
\d b_{mn}={\d a\over \sqrt{-(\partial a)^2}}
[{\cal H}^*_{mn}-{1\over{\cal N}^2}H_{mnp}
({g} ^{ps}+{e^{2\Phi}{\cal{F}}^p{\cal{F}}^s\over
{1-e^{2\Phi}{\cal{F}}^2}}){\partial_s a\over \sqrt{-(\partial a)^2}}]
\end{equation}
and the self-duality
equation for the NS5-brane gauge field $b_2$
\begin{equation}\plabel{sdNS5}
{\cal H}^*_{mn}={1\over {\cal N}^2}H_{mnp}
({g} ^{ps}+{e^{2\Phi}{\cal{F}}^p{\cal{F}}^s\over
{1-e^{2\Phi}{\cal{F}}^2}}){\partial_s a\over \sqrt{-(\partial a)^2}},
\end{equation}
with ${\cal N}$ and ${\cal H}^*_{mn}$ defined in \p{r} and
\p{Gmn}.

In addition to the worldvolume diffeomorphisms and the symmetries \p{symns2}
and \p{symns21}, the action \p{ns2ac} (by construction) has gauge symmetries
(\ref{gb}) and (\ref{u1}).

Thus, we have obtained the action describing the worldvolume dynamics of
the bosonic 5-brane propagating in the `Kaluza--Klein' part (\ref{4}) of the
IIA $D=10$ supergravity background.
In the next section we extend this action to describe coupling
of the NS5--brane to antisymmetric gauge fields of IIA $D=10$ supergravity.

\section{Coupling to the background gauge fields}

When the M5-brane couples to the 3-form background field
$\hat{C}^{(3)}$  of $D=11$ supergravity
the field strength $H_3$ gets extended by the worldvolume
pullback of $\hat{C}^{(3)}$
\begin{equation}\plabel{h3c}
H^{(3)}\rightarrow \hat{H}^{(3)} = db^{(2)} - \hat{C}^{(3)}.
\end{equation}
As a result, up to a total derivative, the
variation \p{dL} of the action \p{ac} with respect to
$b_{mn}(\xi)$ and $a(\xi)$ acquires an additional term in comparison
with eq. \p{dL0}
\begin{equation}\plabel{dL01}
\int\, d\xi^6 \d {\cal{L}} = -
\int\, \left[d(da \wedge {\cal F}_2)  \wedge
(\d b_2
- \d a {\cal F}_2) + {1 \over 2} d\hat{C}_3 \wedge \d b_2\right ],
\end{equation}
The symmetries \p{sym} and  \p{sym1} spoiled by the
last term of (\ref{dL01}) are restored if to the action \p{ac} one adds
the Wess-Zumino term \cite{aharony}
\begin{equation}\plabel{wzM5}
S_{WZ}= \int_{{\cal M}^6}\,( \hat{C}^{(6)}+{1\over 2}d b^{(2)}\wedge
\hat{C}^{(3)}),
\end{equation}
As it was shown in \cite{5br}, the symmetries  \p{sym} and
\p{sym1} uniquely fix the relative factor between $S_{WZ}$ and
the action \p{ac}.
In (\ref{wzM5}) $\hat{C}^{(6)}$ is the pullback of a six--form gauge
potential whose field strength is $D=11$ Hodge--dual to the field strength
of $\hat{C}^{(3)}$
\begin{equation}\plabel{c6c3}
d\hat{C}^{(6)} + {1 \over 2} \hat{C}^{(3)}\wedge  d\hat{C}^{(3)} =
{}^*d\hat{C}^{(3)}
\end{equation}
In addition to the symmetries \p{sym}
and \p{sym1} with ${H}^{(3)}$ generalized as in \p{h3c}, the M5-brane
action \p{ac} extended by the Wess--Zumino term  \p{wzM5} is invariant
under the following transformations of the antisymmetric gauge fields
\begin{equation}\plabel{gtr}
\d \hat{C}^{(6)}=d \hat{\varphi}^{(5)} - {1\over 2}\d \hat{C}^{(3)}\wedge
\hat{C}^{(3)},\ \ \ \d \hat{C}^{(3)}=d \hat{\varphi}^{(2)},
\end{equation}
\begin{equation}\plabel{gtrB}
\d b^{(2)}=\hat{\varphi}^{(2)}({\hat X}(\xi)).
\end{equation}

To get the form of the coupling of the NS5--brane to the antisymmetric
gauge fields of type IIA D=10 supergravity we should dimensionally reduce
$\hat{C}^{(3)}$,  $\hat{C}^{(6)}$  and the Wess--Zumino term \p{wzM5}
of the M5--brane. The dimensional reduction of $\hat{C}^{(3)}$ produces
a ten--dimensional R--R three--form ${C}^{(3)}$ and an NS--NS two--form
 $B^{(2)}$
\begin{equation}\plabel{c32}
 \hat{C}^{(3)}
 ={1\over{3!}}d{\hat X}^{\hat{\underline l}} \wedge
 d{\hat X}^{\hat{\underline n}} \wedge
 d{\hat X}^{\hat{\underline m}}
 \hat{C}_{\hat{\underline m}\hat{\underline n}\hat{\underline l}}(\hat{X})
  =
 \end{equation}
 $$
 = {1\over{3!}}d{X}^{\underline l} \wedge d{X}^{{\underline n}}
 \wedge
 d{X}^{{\underline m}}{C}_{{\underline m}{\underline n}{\underline l}}(X)
 +{1\over 2}d{X}^{{\underline n}}\wedge d{X}^{{\underline m}}
B_{\underline m\underline n}(X) \wedge \left(dy-dX^{\underline
l}A_{\underline l}\right) \equiv $$ $$ \equiv {C}^{(3)}+
B^{(2)}\wedge {\cal F},  $$
and the dimensional reduction of
$\hat{C}^{(6)}$ produces a ten--dimensional five--form ${C}^{(5)}$
and a six--form ${B}^{(6)}$ which are dual to ${C}^{(3)}$ and
$B^{(2)}$, respectively,
\begin{equation}\plabel{c65}
\hat{C}^{(6)}=
 B^{(6)}+C^{(5)}\wedge {\cal F}, \end{equation} the duality relations can
 be easely derived by the dimensional reduction of eq. (\ref{c6c3}).

Thus, the field strength of the self-dual gauge field of the NS5--brane
coupled to the $D=10$ background gauge fields is extended as follows
\begin{equation}\plabel{h3ex}
{H}^{(3)}=d b^{(2)}- C^{(3)}-B^{(2)}\wedge {\cal F},
\end{equation}
and the NS5--brane action (\ref{ns2ac}) is enlarged with the following
Wess-Zumino term
\begin{equation}\plabel{ns2wz}
S_{WZ}= \int_{{\cal M}_6}\,\left[ B^{(6)}+C^{(5)}\wedge{\cal F}+
{1\over 2}d b^{(2)}\wedge C^{(3)}+{1\over 2}d b^{(2)}\wedge B^{(2)}\wedge
{\cal F}\right],
\end{equation}
where ${\cal F}=d\xi^m(\partial_m y-A_m)$ is now the worldvolume pullback of
the one--form \p{c32}.

We have now obtained the action for the NS5--brane
propagating in a background of the bosonic sector of type IIA $D=10$
supergravity
$$
S=-\int\,d^6\xi\,e^{-2\Phi}\sqrt{-\det({g} _{mn}-e^{2\Phi}{\cal{F}}_m
{\cal{F}}_n)}\sqrt{\det\left({\d_m}^n+i{e^{\Phi}({g} _{mp}-
e^{2\Phi}{\cal{F}}_m
{\cal{F}}_p)\over {{\cal N}\sqrt{\det({\d_m}^n-e^{2\Phi}{\cal{F}}_m
{\cal{F}}^n)}}}{H}^{*np}\right)}
$$
$$
-{1\over 4} \int d^6\xi\, \sqrt{-{g} }
{1 \over{\cal N}^2}
{H}^{*mn}H_{mnk}\left({g} ^{kp}+{e^{2\Phi}{\cal{F}}^k{\cal{F}}^p
\over{1-e^{2\Phi}{\cal{F}}^2}}\right){\partial_p a \over \sqrt{-(\partial
a)^2}}
$$
\begin{equation}\plabel{NSIIA}
+ \int_{{\cal M}_6}\, \left( B^{(6)}+C^{(5)}\wedge {\cal F}+
{1\over 2}d b^{(2)}\wedge C^{(3)}+{1\over 2}d b^{(2)}\wedge B^{(2)}\wedge
{\cal F}\right).
\end{equation}

This action is invariant under the worldvolume gauge
transformations \p{gb}, \p{symns2}, \p{symns21} and \p{u1}, with
$H^{(3)}$ now having the form (\ref{h3ex}), and under
target--space gauge transformations
$$ \d C^{(3)}= d\varphi^{(2)}+
d\varphi^{(1)} \wedge A,\qquad \d B^{(2)}=d\varphi^{(1)},\qquad \d
A=d\varphi^{(0)},\qquad $$
\begin{equation}\plabel{gtr2a}
\d b^{(2)}=\varphi^{(2)}-\varphi^{(1)}\wedge dy,\qquad
\d y=\varphi^{(0)},
\end{equation}
under which $\d {H}^{(3)}=0$, and
$$
\d B^{(6)}=d\varphi^{(5)}
+ d\varphi^{(4)} \wedge A
-{1\over 2}d\varphi^{(2)}\wedge C^{(3)}
-{1\over 2}d\varphi^{(1)}\wedge  A \wedge C^{(3)},
$$
\begin{equation}\plabel{gtr2a1}
\d C^{(5)}=d\varphi^{(4)}-{1\over 2}d\varphi^{(2)}\wedge B^{(2)}
+{1\over 2}d\varphi^{(1)}\wedge C^{(3)}
-{1\over 2}d\varphi^{(1)}\wedge A \wedge B^{(2)}.
\end{equation}

Before proceeding with the consideration of the full super--NS5--brane action
let us demonstrate how the action of ref. \cite{BLO,lozano} is obtained from
eq. (\ref{NSIIA}).

\section{NS5--brane action in the second order approximation.}

The action of \cite{BLO,lozano} is a second--order approximation in powers of
$H_{mnk}$ of the NS5--brane action, with the self-duality condition
being regarded as an extra (actually on--shell) constraint.
To get the second--order action we should
expand  (\ref{NSIIA}) in series of  $H^{(3)}$ and truncate it down to the
second order in $H^{(3)}$ assuming the worldvolume gauge field to be weak.
Since the Wess--Zumino term is already linear
and quadratic in $H$, we shall write down only
the ``kinetic'' part of the action
\footnote{We should note that our choice of the dimensionally reduced
$C_3$ and $C_5$ differs from that in \cite{BLO,lozano},
so the Wess-Zumino term in Eq. (\ref{NSIIA}) is related to the one
of Refs. \cite{BLO,lozano} by the following field redefinition:
$$
y\rightarrow c^{(0)}, \qquad
b^{(2)}\rightarrow a^{(2)},
$$
$$
B^{(2)}\rightarrow B^{(2)},\qquad
C^{(3)}-B^{(2)}\wedge A \rightarrow C^{(3)},
$$
$$
C^{(5)}\rightarrow C^{(5)}-{1\over 2}C^{(3)}\wedge B^{(2)}, \qquad
B^{(6)}-C^{(5)}\wedge A \rightarrow -{\tilde B}^{(6)}.
$$
The WZ term of \cite{BLO,lozano} also contains the curl of an
auxiliary worldvolume 5-form field which ensures the exact gauge invariance of the WZ term.
}.

To carry out such a truncation the simplest way is to first truncate
the M5--brane action \p{ac} and then perform its dimensional reduction.
Up to the second order in $H^{(3)}$ the M5--brane action has the form
\begin{equation}\plabel{expan}
S=-\int\,d^6\xi\, \sqrt{-\hat{g}}\left[1-{1\over 4}\hat{H}^*_{mn}\hat{H}^{*mn}+
{1\over 4}\hat{{H}}^{*mn}H_{mnp}\hat{v}^p+ \dots\right]
\end{equation}
with the self-duality condition \p{selfdmn} reducing to
\begin{equation}\plabel{sdcond1}
H_{mnl}-\hat{H}^{*}_{mnl}=0.
\end{equation}

Taking into account the expressions
$$
\hat{H}^{*}_{mnk}={1\over 3!\sqrt{-\hat{g}}}\e_{mnkqrs}H^{qrs}
$$
and
$$
\e_{mnlpqr}\e^{mnlstv}=-(3!)^2 \d_{p}^{[s}\d_{q}^{t}\d_{r}^{v]},
$$
after some algebra one can rewrite \p{expan}
in the following form \cite{pst}
\begin{equation}\plabel{expan1}
S=-\int\,d^6\xi\, \sqrt{-\hat{g}}\left [1-
{1\over 24} H_{mnl} H^{mnl}-
{1\over{8{\widehat{\partial a\partial a}}}}
\partial_m a( H^{mnl}-\hat{H}^{*mnl})
(H_{nlp}-\hat{H}^*_{nlp})\partial^p a
+\dots\right].
\end{equation}

Discarding in (\ref{expan1}) the term containing the auxiliary field
$a(\xi)$ and the anti-selfdual tensor $ H-\hat{H}^{*}$
(which is zero on the mass shell \p{sdcond1}), and carrying out the direct
dimensional reduction of (\ref{expan1}) we recover the NS5--brane action of
\cite{BLO,lozano}
$$
S=-\int\,d^6\xi\,
e^{-2\Phi}\sqrt{-\det({g}_{mn}-e^{2\Phi}{\cal{F}}_m
{\cal{F}}_n)}
\left [1-
{1\over 24}(e^{2\Phi}H_{mnk}H^{mnk} \right.
$$
\begin{equation}\plabel{expan2}
\left. +3{e^{4\Phi}\over
{1-e^{2\Phi}{\cal{F}}^2}}{\cal{F}}_m H^{mnk} H_{nkp}{\cal{F}}^p)+\dots \right].
\end{equation}
Alternatively, the action (\ref{expan2}) can be obtained directly by truncating
the NS5--brane action (\ref{NSIIA}), and discarding terms containing the
auxiliary field and the linearized NS5--brane self--duality condition \p{sdNS5}
$$
{H}^{*}_{mnl}-
H_{mnp}
\left({\delta}^p_l+{e^{2\Phi}{\cal{F}}^p{\cal{F}}_l\over
{1-e^{2\Phi}{\cal{F}}^2}}\right)=0.
$$

\section{The $\kappa$--symmetric super--NS5--brane action}
To generalize the results of previous sections to  describe the propagation of
an NS5--brane in a curved IIA $D=10$ target superspace parametrized by
ten bosonic coordinates $X^{\underline m}$ and 32--component Majorana-spinor
fermionic coordinates $\Theta^{\underline\alpha}$ forming a IIA, $D=10$
superspace coordinate system
\begin{equation}\plabel{Z}
Z^{\underline M}=(X^{\underline m},\Theta^{\underline\alpha}),
\end{equation}
we again start with
an M5--brane propagating in a generic D=11 supergravity background
parametrized by eleven bosonic coordinates $\hat X^{\hat{\underline m}}$
and  32--component Majorana-spinor
fermionic coordinates $\Theta^{\underline\alpha}$ forming a $D=11$
superspace coordinate system
\begin{equation}\plabel{hatZ}
\hat Z^{\hat{\underline M}}=
(\hat X^{\hat{\underline m}},\Theta^{\underline\alpha})=(Z^{\underline M},y),
\end{equation}
where we have separated the eleventh coordinate $y=X^{10}$ keeping in mind
the dimensional reduction of $D=11$ superspace down to
type IIA $D=10$ superspace.

$D=11$ superspace geometry is described by a supervielbein
\begin{equation}\plabel{hatE}
\hat E^{\hat{\underline A}}=
d\hat Z^{\hat{\underline M}}
\hat{E}^{~~\hat{\underline A}}_{\hat{\underline M}} (\hat{Z}) =
(\hat E^{\hat{\underline a}}, \hat E^{{\underline\alpha}}),
\end{equation}
where $\hat{\underline A}=(\hat{\underline a},\underline\alpha)$
are locally flat tangent superspace indices, by a superconnection
\begin{equation}\plabel{hatOm}
\hat w_{\hat{\underline A}}^{~~\hat{\underline B}}=
d\hat Z^{\hat{\underline M}}
\hat w_{\hat{\underline M}}{}_{\hat{\underline A}}^{~~\hat{\underline B}}
(\hat{Z}),
\end{equation}
and by a three--superform generalization of the bosonic gauge field \p{c32}
\begin{equation}\plabel{hatC}
\hat C^{(3)}={1\over{3!}}d\hat Z^{\hat{\underline N}}  \wedge
d\hat Z^{\hat{\underline M}} \wedge
d\hat Z^{\hat{\underline L}}C_{\underline{\hat L\hat M\hat N}}
(\hat{Z}).
\end{equation}

The supervielbein, the superconnection and the gauge superfield
are subject to supergravity constraints which put the superfield
formulation of eleven--dimensional supergravity on the mass shell.
An explicit form of the $D=11$ supergravity constraints relevant
to the description of M5--brane dynamics the reader may find in
\cite{cl,M5,ckp,physrep}.

The super--M5--brane action has the similar form as the bosonic action \p{ac}
enlarged with the WZ term \p{wzM5}, where the worldvolume induced
metric is now
\begin{equation}\plabel{sind}
\hat g_{mn}=\partial_m\hat Z^{\hat{\underline M}}\partial_n
\hat Z^{\hat{\underline N}}
\hat E^{~\hat{\underline a}}_{\hat{\underline N}}
(\hat{Z})
\hat E_{\hat{\underline M}\hat{\underline a}}
(\hat{Z}),
\end{equation}
and $\hat{C}^{(3)}$ and $\hat{C}^{(6)}$ are worldvolume pullbacks of the
three--superform \p{hatC} and its six--superform
dual \cite{cl,bbs}.

In addition to all symmetries discussed above and target--space
superdiffeomorphisms the super--M5--brane action is invariant under the
following fermionic $\kappa$--symmetry transformations \cite{M5,M51}
\begin{equation}\plabel{kappa}
 i_\kappa \hat{E}^{\underline{\hat{a}}}
 \equiv
 \delta_\kappa \hat{Z}^{\underline{\hat{M}}}
 \hat{E}_{\underline{\hat{M}}}^{~\underline{\hat{a}}}(\hat{Z}) = 0,
 \qquad
 i_\kappa \hat{E}^{\underline{\hat{\a}}}
 \equiv
 \delta_\kappa \hat{Z}^{\underline{\hat{M}}}
 \hat{E}_{\underline{\hat{M}}}^{~\underline{\hat{\a}}}(\hat{Z}) =
 (I- \bar{\Gamma})^{\underline{\hat{\a}}\underline{\hat{\b}}}
 \kappa_{\underline{\hat{\b}}},
\end{equation}
 $$
 \delta_\kappa b_2 = i_\kappa \hat{C}_3 \equiv
 {1 \over 2}
 d\hat{Z}^{\underline{\hat{M}}_3} \wedge
 d\hat{Z}^{\underline{\hat{M}}_2} ~
 \delta_\kappa \hat{Z}^{\underline{\hat{M}}_1}
 \hat{C}_{\underline{\hat{M}}_1\underline{\hat{M}}_2\underline{\hat{M}}_3}
 (\hat{Z}), \qquad
 \delta_\kappa a = 0.
$$
where the spinor matrix $\bar\Gamma$ has the following expansion
in products of $D=11$ Dirac matrices
\begin{equation}\plabel{barG}
\bar{\Gamma} = { \sqrt{-\hat{g}}\over
\sqrt{-det(\hat{g}+ i\hat{H}^*)}}
\left( \hat{\Gamma}^{(6)} + {i \over 2}  \hat{H}^*_{mn}\hat{v}_{l}~
( \hat{\Gamma}^{mnl}) + \hat{t}_m \hat{v}_n ( \hat{\Gamma}^{mn})
\right)
\end{equation}
$$
 \hat{\Gamma}^{(6)} = { 1 \over 6!} \varepsilon^{m_1 \ldots m_6}
 \hat{\Gamma}_{m_1} \ldots
 \hat{\Gamma}_{m_6}, \qquad
 \hat{\Gamma}_{m} \equiv ~  \partial_m \hat{Z}^{\underline{\hat{M}}}
 \hat{E}^{~\underline{\hat{a}}}_{\underline{\hat{M}}} (\hat{Z}) ~
 \hat{\Gamma}_{\underline{a}},
 \qquad
 $$
 $$
 \hat{t}^m = ~
 { 1 \over 8} \varepsilon^{mnkplq}
 ~\hat{H}^*_{nk}
 ~ \hat{H}^*_{pl}~\hat{v}_{q} \equiv
 { 1 \over 8} \varepsilon^{mnkplq}
 ~\hat{{\cal H}}^*_{nk}
  ~\hat{{\cal H}}^*_{pl}~\hat{v}_{q}.
$$
As is characteristic of all superbranes,
for the M5--brane action to be $\kappa$--symmetric the superbackground must
satisfy the supergravity constraints \cite{cl,M5,physrep}. When they
are taken into account, from \p{kappa} we get
\begin{equation}\plabel{kappa1}
 \delta_\kappa \hat{H}_3 = -
 i_\kappa \hat{F}_4=
 \hat{E}^{\underline{\hat{a}}} \wedge
 \hat{E}^{\underline{\hat{b}}} \wedge
 \hat{E}^{\underline{{\a}}}
 \left(
\Gamma_{\underline{\hat{a}}\underline{\hat{b}}}
(I - \bar{\Gamma}) \right)
_{\underline{{\a}}\underline{{\b}}}
\kappa^{\underline{{\b}}},
\end{equation}
$$
 \delta_\kappa \hat{g}_{mn} = -4i
 \hat{E}^{~~\underline{{\a}}}_{(m}
 \left( \Gamma_{n)}
(I - \bar{\Gamma}) \right)
_{\underline{{\a}}\underline{{\b}}}
\kappa^{\underline{{\b}}}.
$$

\bigskip

We now turn to the consideration of the super--NS5--brane action.
It can be obtained from the super--M5--brane action by the direct
dimensional
reduction {\sl of the $D=11$ supergravity superfields}.
A consistent ansatz for the dimensionally reduced supervielbein \p{hatE}
was proposed in \cite{ansatz}.
This is the following superfield
generalization of eq. \p{3}
\begin{equation}\plabel{hatEE}
\hat E^{\underline a}=e^{-{1\over 3}\Phi(Z)}E^{\underline a}, \quad \hat
E^{10}= e^{{2\over 3}\Phi(Z)}(dy-dZ^{\underline M} A_{\underline
M}(Z))\equiv e^{{2\over 3}\Phi(Z)}{\cal F},
\end{equation}
\begin{equation}\plabel{ES}
\hat{E}^{\underline{{\alpha }}}= e^{-{1\over
6}\Phi(Z)}E^{\underline{{\alpha}}}(Z)+{\cal
F}\chi^{\underline{{\alpha}}}(Z),
\end{equation}
where $E^{\underline A}(Z)= dZ^{\underline M}~E_{\underline
M}^{~\underline A}= (E^{\underline a}, E^{\underline\alpha})$ are
supervielbeins of type IIA $D=10$ supergravity, $\Phi(Z)$ is the
dilaton superfield, $A_{\underline M}(Z)$ are components of the
one--form gauge superfield $A=dZ^{\underline M} A_{\underline
M}(Z)$, and $\chi^{\underline\alpha}(Z)$ is a Grassmann--odd
Majorana spinor superfield,  which is actually the Grassmann
derivative of the dilaton superfield $\Phi(Z)$.

The superfields which describe IIA $D=10$ supergravity are subject
to the constraints which are obtained from the $D=11$ supergravity
constraints using the ansatz \p{hatEE},\p{ES} and solving for
Bianchi identities. Different forms of these constraints have been
considered in \cite{gates}, \cite{ansatz,c2,town}.

We do not write the super--NS5--brane action explicitly since it has
exactly the same form as Eq. \p{NSIIA} where now the worldvolume induced
metric is
\begin{equation}\plabel{sind10}
g_{mn}=\partial_mZ^{\underline M}\partial_nZ^{\underline N}
E^{~\underline a}_{\underline N}(Z)E_{\underline M\underline a}(Z),
\end{equation}
and all the bosonic background fields are replaced with
corresponding superfields, in particular, $B^{(6)}$, $C^{(5)}$,
$C^{(3)}$ and $B^{(2)}$ are the worldvolume pullbacks of the type
IIA $D=10$ superforms
\begin{equation}\plabel{10super}
C^{(n)}(Z)={1\over{n!}}E^{\underline A_n}\wedge\dots\wedge
E^{\underline A_1}
C_{\underline A_1\dots\underline A_n}(Z).
\end{equation}
Note that the spinor superfield $\chi^{\underline\alpha}$ does not appear in
the action (\ref{NSIIA}).

The super--NS5--brane action is invariant under $\kappa$--symmetry
transformations obtained from eqs. \p{kappa} by substituting into
the latter the ansatz \p{hatEE} and \p{ES}
\begin{equation}\plabel{kappaNS}
 i_\kappa {E}^{\underline{\hat{\a}}}
 \equiv
 \delta_\kappa {Z}^{\underline{{M}}}
 {E}_{\underline{{M}}}^{~~\underline{{\a}}}({Z}) =
 (I- \bar{\Gamma})^{\underline{{\a}}\underline{{\b}}}
 \kappa_{\underline{{\b}}},
\end{equation}
\begin{equation}\plabel{kappaNS2}
 i_\kappa {E}^{\underline{{a}}} = 0,
 \qquad
 i_\kappa {\cal F} = 0 ~~~\Rightarrow ~~~
 \delta_\kappa y =
 i_\kappa {E}^{\underline{{\a}}}
 A_{\underline{{\a}}} (Z),
\end{equation}
$$
 \delta_\kappa b_2 =
 i_\kappa {C}_3
 + {\cal F} \wedge i_\kappa {B}_2, \qquad
 \delta_\kappa a = 0.
$$ where $A_{\underline{{\alpha}}} (Z)$ is a fermionic component
of the Kaluza--Klein connection form $$ A \equiv
dZ^{\underline{M}} A_{\underline{M}} = E^{\underline{{\a}}}
A_{\underline{{\a}}}+ E^{\underline{a}} A_{\underline{a}}. $$

\section{Conclusion and Discussion}

To summarize, we have obtained the covariant $\kappa$-symmetric
action for the super--NS5--brane in a IIA $D=10$ supergravity
background by the direct dimensional reduction of the M-theory
super-five-brane action. In addition to worldvolume
diffeomorphisms, gauge symmetry, $\kappa$--symmetry and background
supergravity symmetries the super--NS5--brane action possesses
special local symmetries ensuring the covariance of actions with
self-dual gauge fields and serving for deriving the self-duality
condition directly from the action as a consequence of the
equation of motion of the gauge field.


 An interesting problem for future study is
 to construct the Lagrangian description of the consistent coupling
 of a type IIA supergravity action to an NS5-brane source.
The latter requires the construction of a duality--symmetric
version of type IIA supergravity by the dimensional reduction of
the duality-symmetric $D=11$ supergravity \cite{bbs}. The
truncation of such a IIA supergravity action shall produce the
duality--symmetric version of the $N=1$, $D=10$ supergravity,
which should naturally couple to a heterotic five-brane \cite{H5}.
Note that recent investigations of interacting brane actions
\cite{BK} may provide one with a possibility of making this coupling
supersymmetric.

Another problem for further studying is to perform the T-duality
transformation of the complete NS5--brane action and to arrive at
a non--linear and supersymmetric action for a type IIB D=10
Kaluza-Klein (KK) monopole. A quadratic approximation for the
bosonic part of this action has been constructed in \cite{lozano}.
One of possible ways of deriving appropriate T--duality
transformation rules for the antisymmetric gauge fields is to
T--dualize the duality--symmetric version of type IIA supergravity
to the duality--symmetric version of type IIB supergravity
\cite{IIB}.

As it was noted in \cite{hull} and
proved in the second order approximation in \cite{lozano}, the
type IIB D=10 KK monopole is expected to be a self--dual object
under the S--duality symmetry of type IIB supergravity. The
construction of the complete action for the type IIB KK monopole
should allow one to explicitly verify this statement.

\vspace{1cm}
{{\bf Acknowledgements.}} The authors are grateful to
Kurt Lechner, Paolo Pasti and Mario Tonin for interest to this
work and valuable discussions and to Christopher Hull,
Jeffrey Harvey,
Bernard Julia and Kellog Stelle for useful comments. I.B. and A.N.
also acknowledge kind hospitality extended to them at
the Abdus Salam International Centre for Theoretical Physics where part of
this work was done. This work has been partially supported by the
Ukrainian GKNT Grant {\bf 2.5.1/52} and INTAS Grant No 96-0308.

\end{document}